\documentclass[conference]{IEEEtran}
\IEEEoverridecommandlockouts
\usepackage{cite}
\usepackage{amsmath,amssymb,amsfonts}
\usepackage{algorithmic}
\usepackage{graphicx}
\usepackage{textcomp}
\usepackage{xcolor}
\usepackage{multirow}

\def\BibTeX{{\rm B\kern-.05em{\sc i\kern-.025em b}\kern-.08em
    T\kern-.1667em\lower.7ex\hbox{E}\kern-.125emX}}
\begin{document}

\title{UMLsec4Edge: Extending UMLsec to model data protection related requirements in edge computing
\thanks{Work partially funded by the European Union’s Horizon 2020 research and
innovation programme under grant agreement no. 871525 (FogProtect). Useful
discussions with project partners are gratefully acknowledged.}
}


\author{\IEEEauthorblockN{Sven Smolka\textsuperscript{1}, Jan Laufer\textsuperscript{1}, Zoltán Ádám Mann\textsuperscript{2}, Klaus Pohl\textsuperscript{1}}
\IEEEauthorblockA{\textsuperscript{1}\textit{paluno – The Ruhr Institute for Software Technology, University of Duisburg-Essen, Essen, Germany}
}
\IEEEauthorblockA{\textsuperscript{2}\textit{Complex Cyber Infrastructure (CCI) group, University of Amsterdam, Amsterdam, The Netherlands}
}
}

\maketitle

\begin{abstract}
Edge computing enables the processing of data -- frequently personal data -- at the edge of the network.
For personal data, legislation such as the European General Data Protection Regulation requires data protection \textit{by design}.
Hence, data protection has to be accounted for in the design of edge computing systems whenever personal data is involved.
This leads to specific requirements for modeling the architecture of edge computing systems, e.g., representation of data and network properties.

To the best of our knowledge, no existing modeling language fulfils all these requirements.
In our previous work we showed that the commonly used UML profile UMLsec fulfils some of these requirements, and can thus serve as a starting point.

The aim of this paper is to create a modeling language which meets all requirements concerning the design of the architecture of edge computing systems accounting for data protection.
Thus, we extend UMLsec to satisfy all requirements.
We call the resulting UML profile UMLsec4Edge.
We follow a systematic approach to develop UMLsec4Edge.
We apply UMLsec4Edge to real-world use cases from different domains, and create appropriate deployment diagrams and class diagrams.
These diagrams show UMLsec4Edge is capable of meeting the requirements. 
\end{abstract}


\section{Introduction}
\label{sec:intro} 
\textbf{Motivation:} Edge computing enables real-time data processing with low network communication latency and satisfactory quality of service at the same time. Due to these positive characteristics, edge computing systems become increasingly popular \cite{yousefpourAllOneNeedsToKnow}.
So-called edge nodes can perform tasks that most end devices (e.g. IoT devices) are not capable of due to insufficient computing power \cite{yiASurveyOfFogComputing}.
Edge nodes can also pre-process data, for example to reduce the amount of data that is sent to the cloud for further processing \cite{yousefpourAllOneNeedsToKnow}. 
However, processing (personal) data at the edge of the network leads to new challenges. 
For example, edge nodes may use different types of communication (e.g., 5G or WLAN) which offer different levels of data protection.
Regulations such as the General Data Protection Regulation (GDPR) of the European Union \cite{regulation2016regulation} prescribe the protection of personal data.
In particular, the GDPR stipulates the need for data protection \textit{by design}. 
To ensure data protection \textit{by design} in edge computing systems, requirements for modeling the architecture of edge computing systems emerge \cite{laufermodeling} which can be sorted into categories: ``network properties'', ``devices'', ``actors'', and ``data properties''.

\textbf{Problem statement \& Aim of the paper:} Existing modeling languages used in system design focus either on security (e.g. \cite{jurjens2002umlsec, sysmlsec}), or on privacy (e.g. \cite{privuml, PrivacyUMLprofile}), or none of the above.
Although the GDPR requires appropriate security under Article 32, the definition of data protection goes beyond that of security and privacy.
Existing modeling languages cannot capture all aspects of edge computing systems related to data protection, for example, the different levels of data protection stemming from different types of communication technologies \cite{laufermodeling}.  
To the best of our knowledge, there is no modeling language which covers all requirements on the modeling of the architecture of data-protection-compliant edge computing systems.
We identified UMLsec (an extension of the well-known modeling language UML) as a promising starting point to model data protection concerns in edge computing systems, because UMLsec is capable of modeling information security aspects for software systems. 
UMLsec addresses the ``data properties'' requirement category as well as parts of the categories ``network properties'' and ``devices''.

The aim of this paper is to create a modeling language which satisfies the identified requirements concerning the design of the architecture of edge computing systems, accounting for data protection.

\textbf{Contribution \& Approach:} 
To achieve this goal, this paper provides the following contributions.
\begin{itemize}
    \item A new modeling language called UMLsec4Edge that supports the modeling of data-protection-compliant edge computing systems. UMLsec4Edge is an extension of UMLsec, focusing on deployment and class diagrams.
    We focus on these diagram types because the architecture of systems and dependencies between system components have great impact on data protection \textit{by design}.
    \item A formalization of UMLsec4Edge as well as an UML profile created with the modeling platform Papyrus.
    \item Deployment and class diagrams of three real-world use cases which employ edge computing in diverse application domains. The use cases originate from the EU-funded research project FogProtect \cite{ayed2020fogprotect}.
    The diagrams confirm that UMLsec4Edge meets the requirements by addressing the shortcomings of UMLsec.
\end{itemize}
The creation of UMLsec4Edge follows a systematic approach based on the work of Lagarde et al. \cite{lagarde2008leveraging}, including a literature search focusing on UML profiles covering security, privacy, or data protection to assess alternative extension options.

\textbf{Outline:} Sec. \ref{sec:preliminaries} defines terminology and introduces the requirements for modeling the architecture of edge computing systems accounting for data protection.
Sec. \ref{sec:umlsec4edge} describes our systematic development approach and as a result the UMLsec4Edge profile.
Sec. \ref{sec:discussion} shows extracts from UMLsec4Edge diagrams and discusses threats to validity.
Sec. \ref{sec:relatedWork} examines related work, while Sec. \ref{sec:conclusion} concludes the paper.
Additional material, including the full UMLsec4Edge profile, complete diagrams, and further information on the systematic approach and literature review, can be found online\footnote{See https://git.uni-due.de/fogprotect/umlsec4edge}.

\section{Requirements to model data protection in edge computing systems}
\label{sec:preliminaries}

\subsection{Data protection in edge computing systems}

\begin{table*}
\caption{Restrictions of UMLsec resulting in non-fulfilment of the requirements, as well as solutions provided by UMLsec4Edge}
\centering
\begin{tabular}{|c|l|l|l|} 
\hline
\multirow{2}{*}{\begin{tabular}[c]{@{}c@{}}Data protection-\\related requirement\end{tabular}} & \multicolumn{1}{c|}{\multirow{2}{*}{\begin{tabular}[c]{@{}c@{}}Restriction of UMLsec leading to non-fulfilment\\of the requirement\end{tabular}}}            & \multicolumn{2}{c|}{Solution provided by UMLsec4Edge}                                                                         \\ 
\cline{3-4}
                                                                                                & \multicolumn{1}{c|}{}                                                 & \multicolumn{1}{c|}{Stereotype}                                                                  & \multicolumn{1}{c|}{Tag}  \\ 
\hline
\textbf{(R-1)}                                                                                             & Lack of ability to model wireless data transmission               & \begin{tabular}[c]{@{}l@{}}\texttt{<<}\texttt{Wireless}\texttt{>>}, \texttt{<<}\texttt{3G}\texttt{>>}, \texttt{<<}\texttt{4G}\texttt{>>},\\ \texttt{<<}\texttt{5G}\texttt{>>}, \texttt{<<}\texttt{RFID}\texttt{>>}, \texttt{<<}\texttt{NFC}\texttt{>>},\\\texttt{<<}\texttt{Bluetooth}\texttt{>>}, \texttt{<<}\texttt{WLAN}\texttt{>>}~ ~\end{tabular}    &                           \\ 
\hline
\multirow{2}{*}{\textbf{(R-2)}}                                                                            & \begin{tabular}[c]{@{}l@{}}Lack of ability to model the threat of unauthorized\\physical access to device types in edge computing\\systems\end{tabular} & \begin{tabular}[c]{@{}l@{}}\texttt{<<}\texttt{ComputingContinuum}-\\\texttt{Device}\texttt{>>}, \texttt{<<}\texttt{EndDevice}\texttt{>>},\\\texttt{<<}\texttt{EdgeNode}\texttt{>>}, \texttt{<<}\texttt{Cloud}\texttt{>>}\end{tabular} &                           \\ 
\cline{2-4}
                                                                                                & \begin{tabular}[c]{@{}l@{}}Lack of ability to model threats between components\\on the same node\end{tabular}  & \texttt{<<}\texttt{internal}\texttt{>>}                                                                                         &                           \\ 
\hline
\multirow{3}{*}{\textbf{(R-3)}}                                                                            & \begin{tabular}[c]{@{}l@{}}Lack of ability to model relationships between actors\\and data\end{tabular}         & \begin{tabular}[c]{@{}l@{}}\texttt{<<}\texttt{Actor}\texttt{>>},\\\texttt{<<}\texttt{DataTraceability}\texttt{>>}\end{tabular}                                                                                 & \begin{tabular}[c]{@{}l@{}}\\\texttt{rights}, \texttt{obligations}\end{tabular}       \\ 
\cline{2-4}
                                                                                                & \begin{tabular}[c]{@{}l@{}}Lack of ability to model trust relationships between\\actors\end{tabular}       & \texttt{<<}\texttt{Actor}\texttt{>>}                                                                                            & \texttt{trusts}                    \\ 
\cline{2-4}
                                                                                                & \begin{tabular}[c]{@{}l@{}}Lack of ability to model actors with multiple\\data-specific roles\end{tabular}                   & \texttt{<<}\texttt{Actor}\texttt{>>}                                                                                            & \texttt{roles}                     \\
\hline
\textbf{(R-4)} & \multicolumn{3}{|c|}{Already covered by UMLsec} \\
\hline
\end{tabular}
\label{tab:modeling_restrictions_and_solutions}
\end{table*}

The term \textit{data protection} refers to the protection of personal data.
Regulations such as the GDPR \cite{regulation2016regulation} prescribe this protection.
The GDPR requires the enforcement of technical and organizational measures to prevent so called \textit{personal data breaches}.
Personal data breaches occur, e.g., whenever an unauthorized actor accesses personal data.
Any system, and therefore also an edge computing system, which processes personal data must ensure the absence of personal data breaches.

Ensuring data protection in edge computing systems faces new challenges compared to more established computing paradigms like cloud computing.
End devices and edge nodes could differ in hardware configuration, preventing the implementation of uniform data protection mechanisms such as hardware enclaves across all devices.  
Moreover, end devices and edge nodes can be deployed almost anywhere.
Thus, they may not be protected by sufficient physical security measures, so there is a threat of attackers physically compromise them.

Since the GDPR requires systems processing personal data to ensure data protection \textit{by design}, such data protection challenges need to be considered when developing an edge computing system, already starting with the architectural design of the system. 
In our previous work \cite{laufermodeling}, we identified four data-protection-related requirements on modeling languages for modeling the architecture of edge computing system:
\newline
\textbf{(R-1) Network properties:} It must be possible to model different communication types between devices in an edge computing system as well as possible threats posed by them.
\newline
\textbf{(R-2) Devices:} It must be possible to model different device types in an edge computing system as well as threats posed by their use.
\newline
\textbf{(R-3) Actors:} It must be possible to model actors within an edge computing system, as well as their trust in each other, their relationship to data, and their data-specific roles.  
\newline
\textbf{(R-4) Data properties:} It must be possible to model data protection requirements specifying whether data must not be disclosed, manipulated, or deleted. 

\subsection{Modeling edge computing architectures with UMLsec}

In our previous work \cite{laufermodeling}, we analyzed to what extent UMLsec \cite{jurjens2002umlsec} supports modeling data protection requirements and threats to data protection during the design of edge computing systems. 
We concluded that UMLsec  provides a reasonable basis for satisfying the identified requirements:

\textbf{(R-1):} UMLsec introduces stereotypes such as \texttt{<<}\texttt{wire}\texttt{>>} or \texttt{<<}\texttt{LAN}\texttt{>>}, which can be used to assign a connection type to a communication path between nodes in deployment diagrams. 
In addition, UMLsec introduces the adversary model, which represents the threat of unauthorized reading, insertion, and deletion of data during data exchange over a communication path of a certain connection type. 
UMLsec is limited in having only stereotypes representing wired connection types.
In edge computing systems, however, data exchange between nodes often takes place wirelessly. 
Consequently, UMLsec only partially addresses \textbf{(R-1)}.

\textbf{(R-2):} UMLsec allows assigning device types to nodes by stereotypes like \texttt{<<}\texttt{POS device}\texttt{>>}.
The adversary model then allows the representation of the threat of unauthorized physical access to these types of devices.
However, there are no stereotypes for device types common in edge computing systems.
Accordingly, the threat of unauthorized physical access to them cannot be modeled in the adversary model.
In addition, UMLsec cannot model threats occurring when data is exchanged between components placed on the same node.  
Overall, UMLsec fulfills a part of \textbf{(R-2)}.

\textbf{(R-3):} UMLsec has no stereotypes or tags to model actors within an edge computing system as well as their trust in each other, their relationship to the data, and their data-specific roles.  
Accordingly, UMLsec does not address \textbf{(R-3)}.

\textbf{(R-4):} UMLsec introduces stereotypes and tags to model the security requirements of data during data exchange in terms of confidentiality, integrity, and availability.
For example, UMLsec introduces the stereotype \texttt{<<}\texttt{secrecy}\texttt{>>}, which can be attached to dependencies between nodes and components in a deployment diagram to define the security requirement of disallowing data to be read by an attacker during data exchange.
In combination with the adversary model (which represents the threats of data transmission over a communication channel with a specific connection type), it is possible to evaluate whether security objectives are met by the system design.
Thus, UMLsec addresses \textbf{(R-4)}.

Table \ref{tab:modeling_restrictions_and_solutions} summarizes the restrictions that make UMLsec not fulfill the requirements.
The table also lists the stereotypes and tags of our UMLsec4Edge profile (presented in the following section) leading to the fulfillment of the requirements. 

\section{UMLsec4Edge}
\label{sec:umlsec4edge}

\subsection{Systematic Approach towards UMLsec4Edge}

\begin{figure*}[t]
\centering
\includegraphics[width=\textwidth]{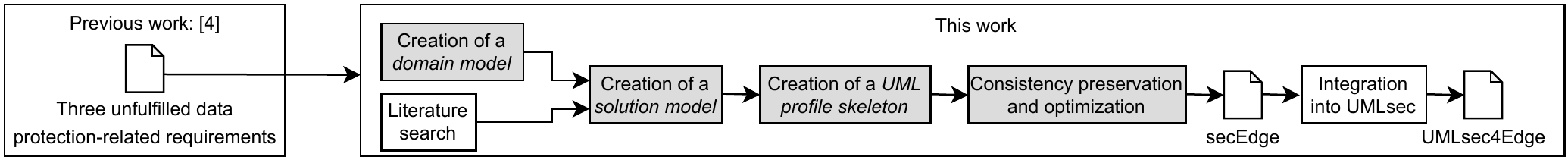}
\caption{UMLsec4Edge creation process (boxes marked in gray indicate a phase according to the approach of Lagarde et al. \cite{lagarde2008leveraging})}
\label{fig:umlsec4edge_creation_process}
\end{figure*}

To create our UMLsec4Edge profile that satisfies all the requirements, we conduct a systematic extension of UMLsec.
Since UMLsec already partially addresses \textbf{(R-1)} and \textbf{(R-2)} in deployment diagrams, it is reasonable to extend UMLsec with respect to deployment diagrams to fully address the requirements.
As UMLsec does not address \textbf{(R-3)}, we extend UMLsec with respect to class diagrams, since the data of a system is often modeled in class diagrams.
Fig. \ref{fig:umlsec4edge_creation_process} shows our systematic approach of extending UMLsec.
We first create a temporary UML profile (called secEdge) following the systematic approach for creating UML profiles according to Lagarde et al. \cite{lagarde2008leveraging}.
Their systematic approach has already been successfully applied by other authors \cite{zoughbi2011modeling, stallbaum2010toward}.
In the first three phases a \textit{domain model}, \textit{solution model} and \textit{UML profile skeleton} are created, and in the last phase consistency preservation and optimization are performed.

\textbf{Phase 1:} The domain model describes the problem, i.e. the required concepts and their relationships to each other.

\textbf{Phase 2:} The solution model is created based on the understanding that emerges from the domain model and contains technical solutions to the problem.
To create the solution model, we also conducted a systematic literature search with the goal to assess different existing solutions.
Hence, we investigated existing UML profiles regarding security, privacy, and data protection.
The literature search was conducted on 23 September 2021 in the Scopus database using the following search string: TITLE-ABS-KEY ( ``UML profile'' AND ( ``data protection'' OR ``security'' OR ``privacy'' ) ).
The search resulted in 129 papers, which we systematically filtered for relevance by applying predefined exclusion and inclusion criteria. 
Included were all papers which contain a UML profile with references to security, privacy, or data protection.
Excluded were all findings which are proceedings, which have not been undergone a peer review process, which are not written in German or English, which do not reference a UML profile, and which do not present UMLsec.
In the end, 47 papers remained, which we analyzed in detail.

\begin{figure}
\centering
\includegraphics[width=\columnwidth]{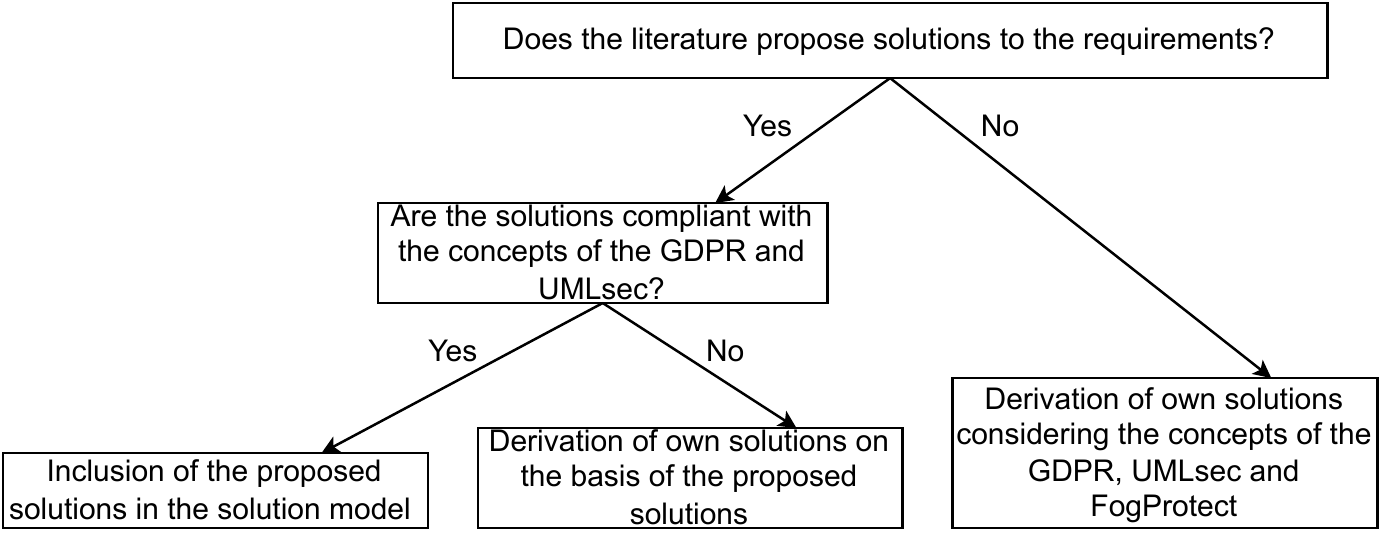}
\caption{Procedure for the creation of the solution model}
\label{fig:decision_tree}
\end{figure}

The decision tree in Fig. \ref{fig:decision_tree} shows how we proceeded in creating the solution model. 
We first investigated whether the literature provides solutions to the requirements.
If the literature provides solutions to the requirements, we reviewed to what extent these solutions are compliant with the concepts of the GDPR and UMLsec. 
If the proposed solutions are compliant with the GDPR and UMLsec, we have included them in the solution model - otherwise, we have derived our own solutions from the proposed solutions taking into account the concepts of the GDPR and UMLsec.
In case the literature does not provide solutions for the requirements, we derived our own solutions considering the concepts of the GDPR, UMLsec and the EU-funded research project FogProtect. 

\textbf{Phase 3:} After the solution model is created, the UML profile skeleton is constructed.
This is done by converting all classes of the solution model to stereotypes in the skeleton, as well as by adding metaclasses to the skeleton to which the stereotypes are mapped. 

\textbf{Phase 4:} After the creation of the skeleton, a validation of the consistency of the profile on the UML meta level is performed as well as an optimization of the profile, i.e. a reduction of the number of stereotypes.

Having systematically created the preliminary secEdge profile, we integrated it with UMLsec, resulting in the final UMLsec4Edge profile.
We also formalized the profile in the Papyrus modeling tool. Papyrus enables tool support, precluding modeling errors.
In the following, we present how we realized \textbf{(R-1)}, \textbf{(R-2)} and \textbf{(R-3)} in UMLsec4Edge and discuss possible alternatives. 
Thereby, we first describe the creation of the solution and subsequently explain where the solution comes from.
In the additional material (see Sec. \ref{sec:intro}) the following artifacts can be found: The models created according to the phases of Lagarde et al. (domain model, solution model, UML profile skeleton, secEdge profile), an extended description on how we realized the requirements, the complete UMLsec4Edge profile, the technical formalization of the profile in Papyrus, and the exclusion and inclusion criteria used in the literature search.

\subsection{Addressing Requirement (R-1)}

In order to fulfill \textbf{(R-1)}, it is necessary to have the means to mark wireless communication channels as such and to model threats posed by such communication channels.
One possibility is to introduce a stereotype such as \texttt{<<}\texttt{Wireless}\texttt{>>}, which represents wireless data transmission between nodes (analogous to the \texttt{<<}\texttt{wire}\texttt{>>} stereotype in UMLsec for a wired data transmission).
Since the communication path between nodes is described by the metaclass ``communicationpath'', the stereotype is mapped to it.

Due to the heterogeneity of devices, several different data transmission types can be used in edge computing systems, posing different security risks. 
Accordingly, it is appropriate to consider such different wireless data transmission types as well.
One possibility would be to assign a tag (for example, of the type of a string or array of an enumeration) to the stereotype \texttt{<<}\texttt{Wireless}\texttt{>>}, which can be used to represent a specific wireless data transmission type.
However, this does not allow the threats of unauthorized data reading, insertion, and deletion, to be mapped to these wireless data transmission types in the adversary model of UMLsec, since threats in the adversary model can only be mapped to stereotypes.

Since the adversary model allows modeling of threats resulting from the use of certain communication types, it is suitable to extend the adversary model since threats must also be represented to fulfill the requirement. 
The extension of the adversary model can be made by representing the wireless data transmission modes as stereotypes.
Thus, an inheritance hierarchy with the stereotype \texttt{<<}\texttt{Wireless}\texttt{>>} as superclass and the stereotypes of the wireless data transmission types as subclasses is appropriate.
Such an inheritance hierarchy can later be extended to include other wireless data transmission types.
Since the UMLsec adversary model is formalized, we have also formally extended it to include the above stereotypes. 
This formal extension can be found in the online material. 

\textbf{Origin of the solution:}
The literature recommends that the following types of wireless data transmission should be considered: 3G, 4G, 5G \cite{mili2019transformation}, Bluetooth \cite{mili2019transformation, decker2009uml, FerranteKM14}, WLAN \cite{mili2019transformation, beckers2012problem, beckers2015supporting, beckers2013problem, mahmoodi2018model}.
Furthermore, based on the case studies of the EU-funded research project FogProtect, we also consider RFID and NFC.
These wireless data transmission types are modeled as an inherited stereotype from \texttt{<<}\texttt{Wireless}\texttt{>>}, allowing the mapping to communication paths.

\subsection{Addressing Requirement (R-2)}

For the realization of \textbf{(R-2)}, it must be possible to represent the threat of unauthorized physical access to device types in an edge computing system. 
Since the adversary model of UMLsec already provides a mapping of this threat to nodes, it is reasonable to extend the adversary model to map the threat to device types used in edge computing systems.
For this purpose, stereotypes must be introduced, which allow to assign such device types to nodes. 
Since various device types could be found in edge computing systems, covering the whole computing continuum from edge to cloud, we introduce a general stereotype \texttt{<<}\texttt{ComputingContinuumDevice}\texttt{>>} to describe a general type of device.  
In order for nodes to be labeled with this stereotype, it must be mapped to the metaclass ``node''.
This stereotype can further act as a superclass and associated subclasses can represent specific device types.

To represent devices in an edge computing system, we introduce stereotypes \texttt{<<}\texttt{EndDevice}\texttt{>>}, \texttt{<<}\texttt{EdgeNode}\texttt{>>}, and \texttt{<<}\texttt{Cloud}\texttt{>>}, which inherit from the stereotype \texttt{<<}\texttt{ComputingContinuumDevice}\texttt{>>}.
Alternatively, the specific device types can be further specified by additional inheritance hierarchies.
For example, \texttt{<<}\texttt{EndDevice}\texttt{>>} could be further refined with subclasses such as ``Smartphone'', ``Smart Vehicle'', or ``CCTV Camera''.
However, since the type of device depends on the scenario in which it is used, this is problematic (e.g., a smartphone could be an end device in one scenario and an edge node in another).
Accordingly, such a further specification would not be appropriate.
Another alternative would be to resolve the inheritance hierarchy into an enumeration.
However, this does not allow the threat of unauthorized physical access to device types to be represented in the adversary model, as this requires stereotypes.

To fulfill \textbf{(R-2)}, it is also necessary to model threats in the data exchange between two components placed on the same node.
The UMLsec adversary model can represent threats in data transmission over a communication path, and thus between components on different nodes.
However, this is not applicable due to the lack of a communication path if the components are placed on the same node.
One solution could be to map data exchange threats in such a case to the node on which the components are placed.
The stereotypes introduced above, which allow the assignment of device types to nodes, could be used for this purpose.
However, this would couple threats to device types, so that labeling a node with the stereotype of a device type would implicitly represent threats in the exchange of data between components on the node.

For a better separation of concerns, such threats should be modeled independently of the device type. 
For this purpose, we introduce a stereotype \texttt{<<}\texttt{internal}\texttt{>>}, which is mapped to the metaclass ``node'' and to which the threats during data transmission between components on the same node can be mapped. Moreover, we formally extended the UMLsec adversary model to include the above stereotypes.  
This formal extension can be found in the additional online material.

\textbf{Origin of the solution:}
From the literature search, it was found that no papers directly address this requirement. 
Thus, the device types in edge computing systems, are based on the labels used in the EU research project FogProtect.
The modeling of threats between components on the same node is based on the adversary model of UMLsec.

\subsection{Addressing Requirement (R-3)}

To fulfill \textbf{(R-3)}, it is necessary to represent relationships between actors and data.
Since data is typically represented as attributes of classes, this requires actors to be represented in class diagrams.
One way to do that is to model actors as classes and introduce a stereotype such as \texttt{<<}\texttt{Actor}\texttt{>>} which makes it apparent that a class represents an actor.
This stereotype is mapped to the metaclass ``class''.
To represent the relationship between actors and data, a stereotype such as \texttt{<<}\texttt{DataTraceability}\texttt{>>} can be attached to the data.
Subsequently, a tag can be introduced for this stereotype, in which the names of the actors are represented, for example, in the form of a string.
Attaching the stereotype to the attributes of a class and modeling the tag, e.g., in the form of a comment outside the class (multiple stereotypes and associated tags result in multiple comments), increases the size of the class diagram, which decreases readability. 

An alternative is to map the stereotype \texttt{<<}\texttt{DataTraceability}\texttt{>>} to the class itself and thus to the metaclass ``class''.
This results in only one tag for all attributes within a class.
Since a class can have multiple attributes and an attribute can have a relationship to multiple actors, the value of this tag must have a well-defined structure.
Such a value can be a string with the following structure: 
The relationship between an attribute of a class and $a$ actors is a (1+$a$)-tuple, which starts with the name of the attribute, followed by the names of the actors’ classes.
Such a tuple is created for each attribute for which a relationship to actors exists.
The value of the string is finally a concatenation of the tuples, where the tuples are separated by a comma.

Since the GDPR provides rights (e.g. the right of a data subject to obtain information about its data) and obligations (e.g. the obligation of a data processor to inform the data subject when collecting personal data) for an actor with respect to data, it is rational to introduce two tags \texttt{rights} and \texttt{obligations}.
These tags must each be formatted according to the string structure above to allow representing the relation between actors and data in terms of rights and obligations.

Furthermore, in order to fulfill \textbf{(R-3)}, it is necessary to model a trust relationship between actors.
For this purpose, the previously introduced stereotype \texttt{<<}\texttt{Actor}\texttt{>>} can be used and extended by another tag like \texttt{trusts}.
This tag may be of type string.
The string could contain the names of the actors (i.e. their class names).
An alternative approach for the type would be a Boolean, which, when taking the value ``true'', states all other actors of the edge computing system trust the actor.
However, this construct is rarely found in reality, since an actor is usually not trusted by all other existing actors \cite{RADAR}.
Actors differentiate between actors whom they (do not) trust.

Finally, in order to meet \textbf{(R-3)}, it must be possible to assign -- possibly multiple -- data-specific roles to actors.
Since roles are tied to actors, it is again a suitable option to extend the stereotype \texttt{<<}\texttt{Actor}\texttt{>>} with another tag, such as \texttt{roles}.
The type of the tag can be a string or an array of an enumeration containing the data-specific roles that actors can take.     
However, this approach can cause syntactical errors in the role assignment when using a modeling tool which does not support checking the contents of strings.
In contrast, an enumeration allows the designer to select data-specific roles from a predefined set, thus preventing syntactical mistakes.

An alternative approach would be to represent the roles in an inheritance hierarchy. 
Here a stereotype, which represents a role in general, acts as superclass and stereotypes which represent concrete roles are associated subclasses.
The stereotype acting as a superclass can be mapped to the metaclass ``class'', whereupon all stereotypes in the inheritance hierarchy can be used to label classes. 
However, in their systematic approach, Lagarde et al. propose a pattern for transforming such an inheritance hierarchy into an enumeration to reduce the number of stereotypes.
Accordingly, we introduce an enumeration called ``RoleType'' which contains the data-specific roles an actor can take.
The type of the tag \texttt{roles} is therefore an array of type ``RoleType''.
In our case, the enumeration includes the roles ``DataSubject'', ``DataController'', ``DataProcessor'' and ``ThirdParty'', since these roles are defined by the GDPR.

\textbf{Origin of the solution:}
Many papers introduce the stereotype \texttt{<<}\texttt{Actor}\texttt{>>} (\cite{alshammari2017uml, bouaziz2014approach, bouaziz2012secure, bouaziz2012applying, neri2013model, salem2012verification, colombo2012towards, cirit2009uml}) to represent actors. 
Since we did not find solutions in the literature for modeling trust relationships between actors and traceability between actors and data, we derive solutions with respect to the GDPR. 
The representation of roles has often been realized in the literature also by an inheritance hierarchy \cite{blanco2015mda, blanco2015architecture, blanco2014showing, triki2010modeling, villarroel2006representing, villarroel2006uml, villarroel2006using, neri2013model, salem2012verification, colombo2012towards, cirit2009uml, wada2006service}.

\section{Evaluation and Discussion}
\label{sec:discussion}

\subsection{Application of UMLsec4Edge to real-world use cases}
\label{subsec:illustration}

To illustrate UMLsec4Edge’s capability of satisfying \textbf{(R-1)}, \textbf{(R-2)}, and \textbf{(R-3)}, we extended the deployment diagrams, class diagrams, and adversary models of three real-world use cases from the domains of Smart Manufacturing, Smart Media, and Smart Cities using UMLsec4Edge.
To ensure representatives of the regarding systems the use cases originate from the EU-funded research project FogProtect.
To ensure reliability we had frequent exchange with the responsible use case owners.
Due to space limitations, we provide excerpts of the diagrams and the adversary model of one of the use cases in this paper. 
The complete extended diagrams and adversary models of the three use cases can be found in the additional material (see Sec. \ref{sec:intro}).
Overall, the use cases represent a basis of an edge computing system.
They consist of cloud centres, edge nodes, and end devices. 
Edge nodes and end devices in particular can exist in a multitude of identical versions up to many thousands of units.
To reduce complexity the use cases are kept to a core that can be scaled up.
However, the core includes all threats to data protection and covers all requirements raised.
Further information can be found online\footnote{see https://fogprotect.eu/results/}.

\textbf{Use Case:}
The Smart Manufacturing use case describes a scenario in which a mobile manufacturing facility called ``Factory in a Box'' (FiaB) \cite{dm2022cost} is installed inside a multimodal transport container.
Inside the container there are end devices such as cameras and robots as well as edge nodes which, e.g., (pre-)process the camera video feed.
FiaB is capable of manufacturing customer-specific orders while in transit.
The inside of FiaB is monitored using cameras.
Authorized personnel can then access monitoring and production data via a terminal (in this case, the personnel is an operator). 
In terms of data-related roles, the customer and authorized personnel are in the role of a data subject. The operator is in the role of both data subject (as he/she may be recorded) and data processor (as he/she can access the customer's personal data and the video recordings).

\begin{figure}
\centering
\includegraphics[width=\columnwidth]{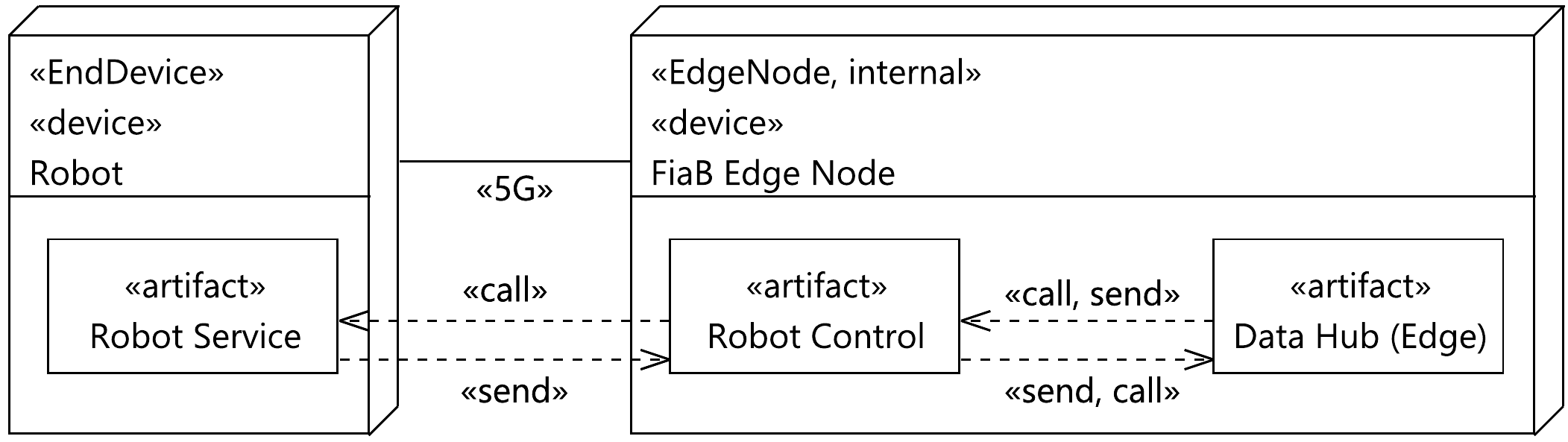}
\caption{Excerpt from the Smart Manufacturing deployment diagram}
\label{fig:excerpt_deployment_diagram_smart_manufacturing}
\end{figure}

\textbf{Example for requirement (R-1):}
Fig. \ref{fig:excerpt_deployment_diagram_smart_manufacturing} shows an excerpt of the Smart Manufacturing deployment diagram extended with the help of the UMLsec4Edge profile. 
Table \ref{tab:excerpt_adversary_model_smart_manufacturing} shows an excerpt of the associated adversary model. 
In the deployment diagram, a robot and an edge node are modeled as nodes which are connected to each other through a communication path. 
One software component (called ``Robot Service'') is placed on the robot and two software components (called ``Robot Control'' and ``Data Hub (Edge)'') are placed on the edge node. 
Software components ``Robot Service'' and ``Robot Control'' communicate with each other through the communication path, while software components ``Robot Control'' and ``Data Hub (Edge)'' communicate with each other within the edge node. 

The UMLsec4Edge profile allows marking the communication path with the stereotype \texttt{<<}\texttt{5G}\texttt{>>}, which enables modeling wireless data transmission between nodes.
In addition, the threats of unauthorized data reading (``read''), data insertion (``insert''), and data deletion (``delete'') can be mapped to the stereotype in the adversary model. 
By enabling the modeling of a wireless data transmission along with the resulting threats, UMLsec4Edge satisfies \textbf{(R-1)}.

\textbf{Example for requirement (R-2):}
In the deployment diagram (see Fig. \ref{fig:excerpt_deployment_diagram_smart_manufacturing}), it is also possible to mark the node ``Robot'' with the stereotype \texttt{<<}\texttt{EndDevice}\texttt{>>} and the node ``FiaB Edge Node'' with the stereotype \texttt{<<}\texttt{EdgeNode}\texttt{>>}.
Furthermore, the threat ``access'' can be mapped to these stereotypes in the adversary model (see Table \ref{tab:excerpt_adversary_model_smart_manufacturing}).
This allows modeling the threat of unauthorized physical access.

Moreover, it is possible to mark the node ``FiaB Edge Node'' with the stereotype \texttt{<<}\texttt{internal}\texttt{>>}.
Hence, the threats ``read'', ``insert'', and ``delete'' can be mapped to this stereotype in the adversary model.
This allows modeling the threat of an attacker being able to read and delete data exchanged between the ``Robot Control'' and the ``Data Hub (Edge)'' as well as insert new data.
By being able to represent both threats of unauthorized physical access to devices and threats in data exchange between components placed on the same device, UMLsec4Edge satisfies \textbf{(R-2)}.

\begin{table}
\small
\centering
\caption{Excerpt from the Smart Manufacturing adversary model}
\begin{tabular}{l|l}
	Stereotype    & Threats  \\ 
	\hline
	\texttt{<{}<}\texttt{5G}\texttt{>{}>}        	   &   \{read, insert, delete\}       \\
	\texttt{<{}<}\texttt{EdgeNode}\texttt{>{}>}     	   &   \{access\}       \\
	\texttt{<{}<}\texttt{EndDevice}\texttt{>{}>}     	   &   \{access\}       \\
	\texttt{<{}<}\texttt{internal}\texttt{>{}>}     	   &   \{read, insert, delete\}       \\
\end{tabular}
\label{tab:excerpt_adversary_model_smart_manufacturing}
\end{table}

\begin{figure}
\centering
\includegraphics[width=\columnwidth]{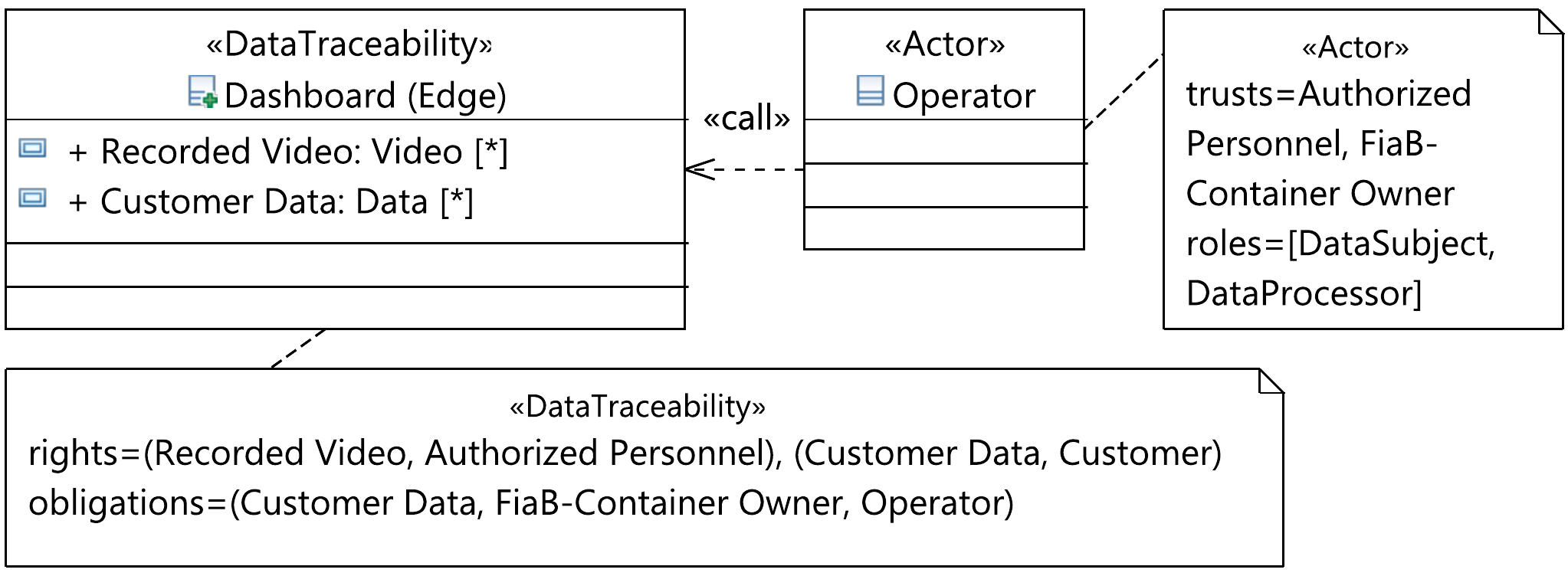}
\caption{Excerpt from the Smart Manufacturing class diagram}
\label{fig:excerpt_class_diagram_smart_manufacturing}
\end{figure}

\textbf{Example for requirement (R-3):}
Fig. \ref{fig:excerpt_class_diagram_smart_manufacturing} shows an excerpt of the Smart Manufacturing class diagram created using the UMLsec4Edge profile.
In the diagram, the class ``Dashboard (Edge)'' represents a software component and the class ``Operator'' represents an actor.
The ``Operator'' accesses the ``Dashboard (Edge)'', which contains the data ``Recorded Video'' and ``Customer Data''. 
With UMLsec4Edge, it is possible to mark the class ``Operator'' with the stereotype \texttt{<<}\texttt{Actor}\texttt{>>} to show this class represents an actor.
Furthermore, it is possible to mark the class ``Dashboard (Edge)'' with the stereotype \texttt{<<}\texttt{DataTraceability}\texttt{>>}.
Subsequently, the values of the tags \texttt{rights} and \texttt{obligations} of the stereotype can be used to define which actor has rights or obligations with respect to which attribute of the class.
The string of tag \texttt{rights} means that actor ``Authorized Personnel'' has rights regarding attribute ``Recorded Video'' and actor ``Customer'' has rights regarding attribute ``Customer Data''.
The string \texttt{obligations} denotes that actors ``FiaB-Container Owner'' and ``Operator'' have obligations towards attribute ``Customer Data''
(Note: Actors ``Authorized Personnel'', ``Customer'' and ``FiaB-Container Owner'' are not presented in the extract of the diagram).
Thus, it is possible to model relationships (specified in terms of rights and obligations) between actors and data.

The tag \texttt{trusts} of the stereotype \texttt{<<}\texttt{Actor}\texttt{>>} is used to model the trust of the actor ``Operator'' towards other actors.
The ``Operator'' trusts the ``Authorized Personnel'' and the ``FiaB Container Owner''.
Correspondingly, trust relationships between actors can be modeled.

In addition, the stereotype \texttt{<<}\texttt{Actor}\texttt{>>} enables the use of the tag \texttt{roles} to assign data-related roles to an actor.
The ``Operator'' is assigned the roles ``DataSubject'' and ``DataProcessor''. 
Accordingly, the operator can simultaneously take on the role of data subject and data processor.
By allowing relations between actors and data, trust relationships between actors, and assignment of multiple data-specific roles to an actor, UMLsec4Edge satisfies \textbf{(R-3)}.

\subsection{Threats to Validity}
\label{subsec:threatstovalidity}

\textbf{Internal validity:}
An internal validity risk exists due to possible syntactic and semantic errors which may have been made when creating the UMLsec4Edge profile.
In order to avoid errors, the creation process of the UMLsec4Edge profile was based on the systematic UML profile creation approach by Lagarde et al. including a consistency preservation of the created UML profile on meta-level \cite{lagarde2008leveraging}.
The usage of the Papyrus modeling tool prevented syntactic and semantic errors.

Furthermore, there may be the risk that the extensions to UMLsec only focus on solving the limitations identified for three use cases in previous work which were co-created by the authors.
To address this risk, we focused on meeting requirements instead of solving limitations. Additionally, we conducted a systematic literature review to find ideas for possible realizations of the requirements.
We investigated 47 papers, about half of which provided fruitful ideas.

\textbf{Reliability:}
A threat to reliability exists for the application of the UMLsec4Edge profile because it was done by the same group of authors who defined the profile.
To minimize the risk we have distributed the creation and the evaluation of UMLsec4Edge within our team.

\textbf{External validity:}
Although UMLsec4Edge fulfills all data protection-related requirements in the three real-world use cases used in our evaluation, there are other use cases in edge computing which have not been considered.
It is therefore not possible to conclude whether the data protection-related requirements are also fulfilled by UMLsec4Edge in all other use cases. 
However, this risk exists with any modeling language, since complete coverage of all possible use cases in edge computing is not possible.

\section{Related work}
\label{sec:relatedWork}
Related work covers several modeling languages which are based on UML and deal with modeling security and / or privacy aspects.
Privacy-focused UML profiles like \cite{basso2018privapp, canovas2018uml, privuml, PrivacyUMLprofile} support the development of privacy-aware applications.
Modeling languages that focus on modeling security aspects \cite{jurjens2002umlsec, escamilla2021iotsecm, sysmlsec} especially focus on confidentiality, integrity, and availability of data.
However, both privacy and security modeling languages lack important contributions in regard to data protection and thus do not address our requirements.

Ahmadian et al. \cite{ahmadian2017model} present two UML profiles regarding data protection.
One of the profiles allows the modeling of personal data including their sensitivity, while the other profile, which is an extension of UMLsec, allows the modeling of access to data based on data-specific roles. 
However, both profiles are only used in UML behavior diagrams.
Thus, their approach is complementary to UMLsec4Edge.
In the future, both approaches could be combined to address both, modelling of system architecture as well as modelling of system behavior.

Modeling languages such as \cite{escamilla2021iotsecm} and \cite{thingMLsec} extend UMLsec to model secure systems in the Internet of Things (IoT).
Ficco et al. also extend UMLsec by introducing new stereotypes to model a secure deployment of cloud applications \cite{cloudbasedUMLsecextension}.
These approaches do not consider modeling IoT and cloud elements as part of an edge computing system.
Thus, they are not capable of meeting our requirements.

When modeling secure business processes, several extensions to the Business Process Model and Notation (BPMN) can be used \cite{agostinelli2019achieving, salnitri2017designing, chergui2018valid}.
While the main focus is on security, some data protection aspects can also be modeled.
However, modeling business processes only partially supports the development of the architecture of (edge computing) systems which take data protection into account.

Approaches like \cite{RADAR} allow to model data protection in cloud systems. Yet, these approaches do not cover characteristics of edge computing, nor are they as extensive as UMLsec. Furthermore, they are not as commonly known and easy to use as a modeling language that is based on the UML.

\section{Conclusions and future work}
\label{sec:conclusion}
This paper addressed the problem of accounting for data protection while modeling the architecture of edge computing systems.
We presented UMLsec4Edge as an extension to UMLsec.
UMLsec4Edge covers deployment and class diagrams.
With the help of UMLsec4Edge, it is possible to model data protection aspects in the design of edge computing systems.
The extensions to UMLsec we have developed represent a coherent construct and complement each other.

We assume UMLsec4Edge can also be applied to related domains such as the Internet of Things and cloud computing.
Further research could verify this hypothesis.
This work can serve as a basis for investigating whether our extensions also contribute to the modeling of data protection aspects in combination with other UMLsec diagram types.
Furthermore, UMLsec4Edge could serve as a communication tool for developers as well as an input for data protection compliance analysis.
An evaluation including domain experts to investigate whether UMLsec4Edge fulfils this purposes is planned as future work.
An automated analysis of UMLsec4Edge diagrams could be implemented by using for example the Object Constraint Language. 
In the future, UMLsec4Edge could serve as a basis to model edge computing systems at runtime to evaluate data protection whenever the system changes.

\bibliographystyle{IEEEtran} 
\bibliography{main}

\end{document}